\documentclass[aps,prx,amsfonts, amssymb, amsmath, reprint, showkeys, nofootinbib, oneside, superscriptaddress]{revtex4-2}

\usepackage{amsthm}
\usepackage{mathtools}
\usepackage{physics}
\usepackage{xcolor}
\usepackage{color}
\usepackage{graphicx}
\usepackage{adjustbox}
\usepackage{setspace}
\usepackage{placeins}
\usepackage[T1]{fontenc}
\usepackage{lipsum}
\usepackage{csquotes}
\usepackage[normalem]{ulem}
\usepackage[pdftex, pdftitle={Article}, pdfauthor={Author}]{hyperref} 
\usepackage{outlines}
\usepackage[ruled,linesnumbered]{algorithm2e}
\usepackage{newtxtext}
\usepackage{newtxmath}

\definecolor{dark-red}{rgb}{0.84,0.15,0.15}
\definecolor{dark-blue}{rgb}{0.15,0.15,0.9}
\definecolor{medium-blue}{rgb}{0,0,0.8}
\definecolor{copper}{rgb}{0.72, 0.45, 0.2}
\definecolor{forest-green}{cmyk}{0.76,0,0.76,0.25}
\definecolor{cerulean}{cmyk}{0.98,0.57,0,0.25}

\usepackage{hyperref}
\hypersetup{
	colorlinks, linkcolor={forest-green},
	citecolor={medium-blue}, urlcolor={cerulean}
}
\usepackage[capitalize,nameinlink]{cleveref}

\begin{document}
	\title{Distributed Quantum Computing via Adaptive Circuit Knitting}
	
	\author{K. Grace Johnson}
	\affiliation{HPE Quantum, Emergent Machine Intelligence, HPE Labs}
	
	\author{Aniello Esposito}
	\affiliation{HPE HPC \& AI EMEA Research Lab, HPE Labs}
	
	\author{Gaurav Gyawali}
	\affiliation{HPE Quantum, Emergent Machine Intelligence, HPE Labs}
	
	\author{Xin Zhan}
	\affiliation{HPE Quantum, Emergent Machine Intelligence, HPE Labs}
	
	\author{Rohit Ganti}
	\affiliation{HPE Quantum, Emergent Machine Intelligence, HPE Labs}
	
	\author{Namit Anand}
	\affiliation{HPE Quantum, Emergent Machine Intelligence, HPE Labs}
	
	\author{Raymond G. Beausoleil}
	\affiliation{HPE Quantum, Emergent Machine Intelligence, HPE Labs}
	
	\author{Masoud Mohseni}
	\thanks{Email: masoud.mohseni@hpe.com}
	\affiliation{HPE Quantum, Emergent Machine Intelligence, HPE Labs}

	\date{\today}

	\begin{abstract}
		Distributing quantum workloads over many Quantum Processing Units (QPUs) is a crucial step in scaling up quantum computers toward practical quantum advantage due to the limitations in size of a single QPU. In the absence of high-fidelity quantum interconnects, circuit knitting could provide a path to computing certain properties of large quantum systems on many QPUs of limited size in a distributed fashion using only classical communication. Circuit knitting partitions large quantum circuits into manageable sub-circuits, however, reconstructing observables in a straightforward manner comes at an exponential cost in sampling and classical post-processing. To mitigate the overhead this technique incurs, we introduce an Adaptive Circuit Knitting (ACK) method that finds efficient partitions of quantum circuits by discovering regions of minimal entanglement between subsystems. We simulate 1D and 2D disordered mixed-field Ising models up to 60 qubits and show that the ACK approach can reduce circuit knitting sampling overheads by up to four orders of magnitude for observables of interest. We highlight our parallel GPU-accelerated implementation and discuss the need for efficient classical simulators to enable distributed quantum algorithm development. Our techniques could enable efficient distribution of quantum simulation for both near-term and fault-tolerant architectures.
	\end{abstract}
	
	\maketitle

	\section{Introduction}
	\label{sec:intro}
	
	Quantum computers have the potential to provide exponential speedups compared to their classical counterparts for a range of applications including simulating quantum systems, quantum machine learning on quantum data, and integer factorization. Instead of replacing classical computers as stand-alone processors, however, quantum computers can be better understood as accelerators or co-processors that can efficiently carry out specialized tasks. As quantum computing is expected to accelerate certain high-performance computing (HPC) workloads, the HPC community has shown considerable interest in strategies for integrating quantum computers into existing and future HPC, forming the emerging field of HPC-QC integration \cite{mohseni2025buildquantumsupercomputerscaling, shehata2025bridging, bravyi2022future}. Hybrid quantum-classical algorithms and application workflow development will be essential for HPC communities and users not only for near-term quantum devices—the Noisy Intermediate-Scale Quantum or NISQ era \cite{preskill2018quantum}—but also for future fault-tolerant quantum computers (FTQCs), as quantum error-correction schemes will rely heavily on classical HPC \cite{mohseni2025buildquantumsupercomputerscaling, caldwell2025platform}. 
	
	It is expected that a central challenge in scaling quantum computers to a sufficient size to achieve utility is distributing execution across many quantum processing units (QPUs) \cite{mohseni2025buildquantumsupercomputerscaling, bravyi2022future}. In the NISQ era, distributing computation across multiple QPUs will be necessary to study systems larger than the processors available today, which are on the order of tens or hundreds of physical qubits. Even in the fault-tolerant era, individual QPUs might not have the required number of qubits to run many important algorithms \cite{mohseni2025buildquantumsupercomputerscaling}. In either case, inter-QPU communication will incur additional performance cost, and partitioning schemes should minimize this cost. Beyond software and algorithmic considerations, distributed quantum computing also emerges naturally from the hardware standpoint, since existing quantum hardware appear to be size-limited. Namely, there is a physical limit to the number of qubits a single QPU can host. For example, although there are some projections that superconducting qubit wafers might host up to $10$k qubits and we could scale this further via wafer-scale integration \cite{mohseni2025buildquantumsupercomputerscaling}, superconducting wirings impose a strict limitation on the total number per QPU \cite{Acharya_2023_multiplexed, Krinner_2019_engineering_cryogenic}. In this case, distributed quantum computing would be the only way to couple such QPUs at a scale where they can tackle nontrivial problems. The hardware-origin of distributed QPUs make these issues essentially agnostic to NISQ or FTQC.
	
	To scale quantum computers and achieve practical high-performance quantum computing, efficiently partitioning and distributing quantum workloads is not only essential in the short term, but may even be optimal in the long term. While distributed fault-tolerant execution will necessitate quantum interconnects---technologies which will require significant development to improve fidelities at scale \cite{CheeWei2025CNOT}---classical communication will also likely have a role to play in logical circuit partitioning, as discussed below. In either case, one has to pay a significant overhead for inter-QPU communication, and efficient partitioning schemes are expected to minimize this cost. Sampling overhead arises because we need to sum over many different disconnected circuits to make up for the entanglement lost due to cutting the gates connecting them. Therefore, to achieve an efficient partitioning, one should minimize the quantum correlations i.e., entanglement between the partitions.
	
	\begin{figure}[t]
		\centerline{\includegraphics[width=0.99\columnwidth]{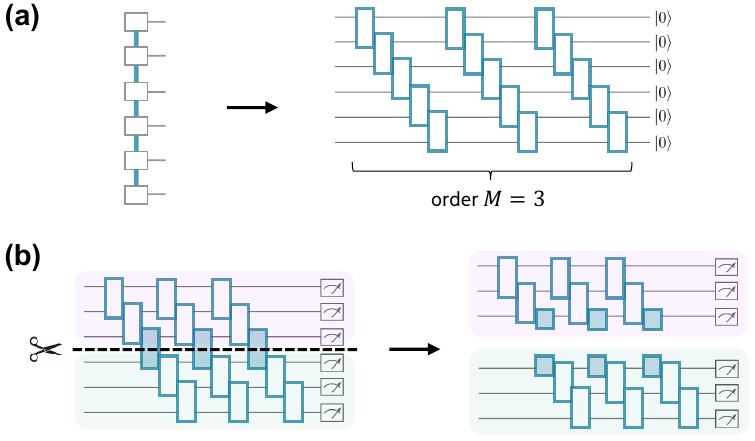}}
		\caption{(a) An MPS (left) with bond dimension $\chi$ (thick lines) can be converted to a quantum circuit tensor network (right) of order $M$. For exact representation, $M=log(\chi)$. (b) Cutting gates in circuit knitting to form two sub-circuits that can be executed independently. After cutting, sub-circuits must be sampled many times to reconstruct (knit) the observable.}
		\label{Fig1}
	\end{figure}
	
	Efficiently partitioning quantum systems is a highly non-trivial task because entanglement can be difficult to characterize. For a given problem it may not be known beforehand which quantum correlations are important to keep (i.e., which qubits should be located on the same QPU) and which are safe to ignore (a good place to partition). Moreover, for dynamical systems, even if one can find low-entanglement boundaries, they may evolve over time in a nontrivial fashion. Nonetheless, such efforts have a long history in quantum science \cite{white1993density, meyer1990multi, manthe2008multilayer}. For the paradigm of gate-based quantum computing prevalent today, the chief concern is partitioning quantum circuits. Circuit knitting has emerged as a promising technique to this end, the primary initial goal being to enable simulating large circuits on devices available today which are limited to tens or hundreds of noisy qubits \cite{bravyi2016trading, peng2020simulating, piveteau2023circuit}. Circuit knitting relies only on classical communication, thus removing dependence on the low-fidelity quantum interconnects available today. The method was originally introduced for error mitigation by using a quasi-probability decomposition to reproduce the output of a large noiseless quantum circuit by execution of many smaller noisy quantum circuits \cite{Temme_quasi_2017,endo_2018_practical_quantum_error_mitigation,piveteau_quasiprobability_2022}. Circuit knitting is of particular interest for hybrid HPC-QC execution because it lends itself to classical parallelization. In circuit knitting, a measured observable is reconstructed by sampling sub-circuits of the original circuit multiple times. Nevertheless, the observable reconstruction comes at an exponential sampling cost. This sampling overhead can be reduced by dynamically optimizing the locations and number of cuts that produce the sub-circuits \cite{piveteau2023circuit}. Although recent work has focused on reducing this exponential overhead \cite{tang2021cutqc, tang2022scaleqc, basu2024fragqc} and even showcased a real-time classical interconnect between two superconducting QPUs \cite{carrera2024combining}, none have used entanglement characterization as a guide to partitioning. In all cases, significant progress is needed to establish the practical advantage of circuit knitting.
	
	In this work, we introduce an \textit{adaptive circuit knitting} (ACK) method for quantum workload distribution and show how it can decrease the sampling overhead of circuit knitting by cutting quantum gates in locations that minimize entanglement between quantum circuit partitions. The paper is outlined as follows. \cref{sec:overview} provides background on the techniques of circuit knitting and quantum circuit tensor networks upon which we build our ACK algorithm, while \cref{sec:ack} describes the ACK method and our parallel, GPU-accelerated implementation. We present our numerical results simulating various 1D and 2D disordered transverse Ising models with longitudinal fields in \cref{sec:numerical_studies}, followed by a discussion in \cref{sec:discussion} and concluding remarks in \cref{sec:conclusion}.

	\section{Overview of tensor networks and circuit knitting}
	\label{sec:overview}
	
	\subsection{Quantum circuit tensor networks}
	\label{subsec:quantum_circuit_tensor_networks}
	
	Tensor networks provide a powerful framework for representing the high-dimensional wavefunctions of quantum many-body systems by decomposing a state vector into a network of local, interconnected tensors. This approach provides a convenient description for low-entanglement states. The most prominent examples of tensor networks include Matrix Product States (MPS) for one-dimensional systems and Projected Entangled Pair States (PEPS) for two-dimensional systems, which naturally capture area-law entangled states \cite{dmrg_2011_schollwock,cirac_2021_mps}. By truncating the internal bond dimensions of these tensors, we can efficiently simulate ground states as well as time evolution using algorithms like the Density Matrix Renormalization Group (DMRG) \cite{white1993density} for systems with a limited amount of entanglement, e.g., ground states of gapped Hamiltonians. 
	
	Tensor network methods have also been used widely for classical simulation of quantum circuits, a primary use case being evaluation of quantum circuit observables via tensor network contraction \cite{berezutskii2025tensor, nguyen2022tensor}. In the same manner as when representing quantum many-body systems, tensor networks can represent quantum circuits with limited entanglement or certain entanglement structures. This raises the interesting prospect of the inverse idea, namely representing a tensor network in a compressed form as a quantum circuit. With this technique, known as quantum circuit tensor networks, a tensor network can be compressed to a circuit with depth only logarithmic in the bond dimension $\chi$ of the original tensor network \cite{lin2021real, haghshenas2022variational}. \cref{Fig1}(a) illustrates this for a simple case of an MPS represented by a quantum circuit with a ladder structure of two-qubit unitary gates. The identity of the two-qubit unitary gates is determined by a variational procedure, with different ans\"{a}tze and optimization methods resulting in varying degrees of expressivity \cite{haghshenas2022variational}. This technique forms the basis for the ACK method and is carried out in a distributed manner on each QPU partition, as discussed in \cref{sec:ack}.

	\subsection{Circuit knitting for distributed quantum computing with classical communication}
	\label{sec:circuit_knitting}
	
	Our ACK method uses circuit knitting to reconstruct a full circuit from circuit partitions, each of which is variationally optimized as noted above. Circuit knitting is built upon the theoretical foundation of quasiprobability decomposition (QPD) \cite{Temme_quasi_2017, endo_2018_practical_quantum_error_mitigation,piveteau_quasiprobability_2022}. QPD allows one to express a unitary channel $\mathcal{U}$ acting across two separate quantum processors $A$ and $B$ as a weighted sum of operations $\mathcal{F}_j$'s, i.e.,
	\begin{equation}
		\mathcal{U} = \sum_j c_j \,\mathcal{F}_j,
	\end{equation}
	where, $\mathcal{F}_j$'s are operations that can be physically realized on the (independent) quantum hardware, for example, local operations (LO), and local operations with classical communication (LOCC). The coefficients $c_j$ are real numbers which can be negative, but we can absorb the sign and normalization factor into the operation $\mathcal{F}_{A/B,j}$ and turn them into a true probability distribution, hence the name quasiprobability. The cost of this sampling is quantified by the 1-norm: 
	\begin{equation}
		\gamma = \sum_j |c_j|, 
		\label{eq:gamma_factor}
	\end{equation}
	the gamma factor. The number of samples required to estimate a linear observable with error $\epsilon$ is $O(\gamma^2/\epsilon^2)$. For two-qubit gates, the gamma factor ranges from $\gamma = 1$ to $\gamma = 7$; for instance, $\gamma(\mathrm{CNOT}) = 3$ and $\gamma(\mathrm{SWAP}) = 7$, while for product unitaries, i.e., $U_{AB} = U_A \otimes U_B$, $\gamma(U) = 1$ \cite{schmitt_2025_cutting_circuits}. \cref{Fig1}(b) shows an example of cutting an order 3 staircase circuit in the middle. Circuit knitting expectation values for this circuit requires QPD of the 3 gates split by the cut, and the sample complexity (the total gamma factor) is the product of individual gamma factors. In this way, the sample complexity of a typical circuit scales \textit{exponentially} i.e., $O(\gamma^{2n}/\epsilon^2)$ in the number of two-qubit gates cut.
	
	Instead of cutting gates by performing QPD on two-qubit gates one by one, we can also perform a \emph{combined cutting}, where a single quasi-probability decomposition is applied jointly across all gates that cross the partition, rather than decomposing each gate independently. Formally, if $\{ U_1, U_2, \dots, U_n \}$ are gates acting across a cut, individual cutting applies QPD to each $U_i$ separately, leading to a scaling like $\prod_j \gamma(U_j)$. In contrast, combined cutting constructs a joint decomposition over the entire set $\{ U_1, \dots, U_n \}$, which can exploit correlations among these gates to reduce the multiplicative growth in the gamma factor.
	As an example, we can start with $n$ Bell pairs that are pre-shared between ancilla qubits on the two partitions. Then, we can use gate teleportation protocols with measurement and feedforward, enabled by two-way classical communication, to apply the desired gates, say $n$ interspersed CNOT gates \cite{piveteau2023circuit}. In this setting, the total number of samples reduces from $O(9^n)$ to $O(4^n)$. More recent protocols do not require classical communication \cite{ufrecht2024jointcutting, schmitt_2025_cutting_circuits}. In Ref. \cite{schmitt_2025_cutting_circuits}, the authors show that for combined cutting of general two-qubit unitaries, the optimal sampling overhead can be achieved even without classical communication. 
	Similarly, Ref. \cite{harrow2025optimalquantum} gives a double-Hadamard test-based protocol for achieving the optimal overhead for cutting a generic unitary without classical communication. However, whether classical communication improves the circuit-knitting overhead for combined cutting of such general unitaries remains an open question.

	In addition to cutting gates, one can also cut wires (time-like cuts) to simulate circuits beyond a device's coherence time \cite{peng2020simulating, lowe2023fast}. Unlike space-like cuts, where the classical communication does not strictly improve the overhead, wire cutting overhead scales strictly multiplicatively without it \cite{brenner2025optimalwirecutting}, whereas communication can significantly improve this via QPD. While utilizing both cut types facilitates distributed computation of generic circuits, the cost is invariably exponential in the number of cuts. Focusing on the former approach, this work targets gate cutting (space-like cuts) and proposes an algorithm to mitigate the exponential sampling overhead for certain quantum systems.
	
	\section{Adaptive circuit knitting}
	\label{sec:ack}
	
	\subsection{Algorithm and theory}
	\label{sec:theory}
	
	\begin{figure*}[ht]
		\centerline{\includegraphics[width=1.9\columnwidth]{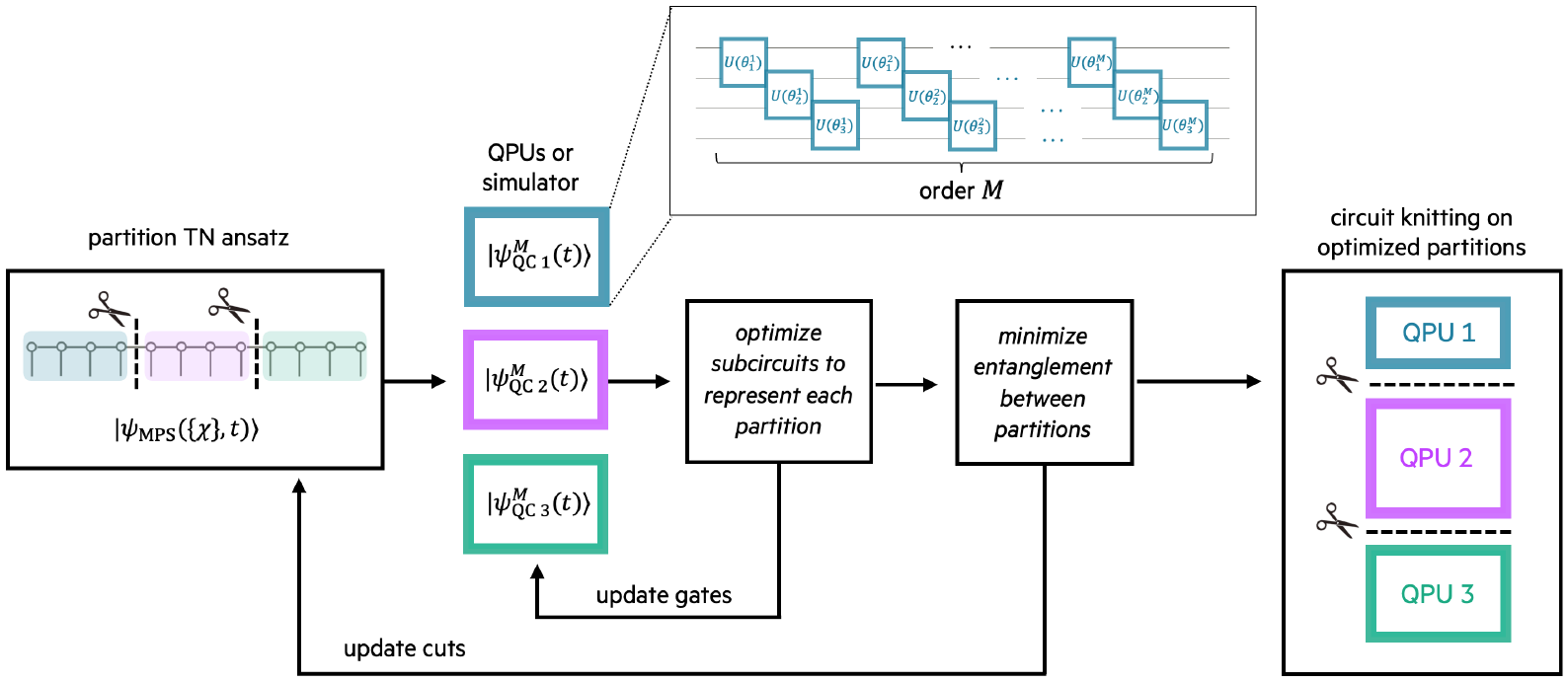}}
		\caption{Schematic of the adaptive circuit knitting method. In the inner loop, a variational optimizer (see work by Lin et al. \cite{lin2021real}) finds two-qubit unitary gate parameters U($\theta$) for partitions of a quantum system based on a tensor network in parallel. In the outer loop, an adaptive procedure finds cuts which minimize entanglement between partitions. After the best cuts are found, observables are reconstructed via circuit knitting.}
		\label{Fig2}
	\end{figure*}
	
	Instead of considering circuit knitting of space-like cuts at the gate-level, we can also think at the level of the states and ask the question---what is the minimum circuit knitting overhead for cutting a density matrix $\rho_{AB}$? It turns out that $\gamma(\rho_{AB})$ is related to a well-studied entanglement monotone called \emph{robustness of entanglement} $R(\rho_{AB})$, first studied in Ref. \cite{vidal1999robustness}, by $\gamma(\rho_{AB})= 1 + 2R(\rho_{AB})$ \cite{piveteau2023circuit}. The robustness of entanglement is an entanglement measure that quantifies the amount of mixing with separable states required to make the given state fully separable. Specifically, it is computed as the closest distance from the set of separable states. Since separable states incur no circuit-knitting overhead, intuitively, it makes sense that the circuit knitting overhead is related to the robustness of entanglement. 
	
	This relationship becomes particularly explicit for pure states. For a pure state with a Schmidt decomposition given by
	\begin{equation}
		\ket{\psi_{AB}} = \sum_j \lambda_j \ket{\psi_{A,j}} \otimes \ket{\psi_{B,j}},
		\label{eq:bipartite_schmidt}
	\end{equation}
	the robustness of entanglement simplifies to a closed form $R(\ket{\psi_{AB}}) = \left(\sum_j \lambda_j\right)^2-1$ \cite{vidal1999robustness}. Consequently, the circuit knitting overhead is quantified by
	\begin{equation}
		\gamma(\ket{\psi_{AB}}) = 2 \left(\sum_j \lambda_j \right)^2 -1.
		\label{eq:gamma-pure-state}
	\end{equation}
	In this form, $\gamma(\ket{\psi_{AB}})$ shares key properties with the entanglement entropy, defined as
	\begin{equation}
		S(\ket{\psi_{AB}}) = -\sum_j \lambda_j^2 \log\lambda_j^2.
	\end{equation}
	Both $\gamma(\ket{\psi_{AB}})$ and $S(\ket{\psi_{AB}})$ are entanglement monotones, invariant under local unitaries, and have the lowest values i.e., 0 or 1 respectively, only when $\ket{\psi_{AB}}$ is unentangled. Thus, in this context, the circuit knitting overhead can be effectively treated as a proxy for entanglement entropy. In particular, they are related to \(\alpha\)-R\'enyi entropies (see \cref{eqn:renyi_definition}) by the following inequality derived in \cref{sec:renyi_entropy}:
	\begin{equation}
		S_{1/2}(\rho) \geq S, \quad  S_{1/2}(\rho) = \log(\frac{\gamma + 1}{2})
		\label{eq:entropy_inequality}
	\end{equation}
	Since the circuit knitting sampling overhead is an entanglement measure, we can achieve a lower sampling overhead by choosing a low-entanglement cut. This is the basis of our \emph{adaptive circuit knitting} (ACK) algorithm. Since ACK depends on the entanglement structure of the quantum state, it is useful to think about the algorithm in the language of tensor networks. The expressiveness of quantum circuit tensor networks allows for a more efficient representation of quantum ground states than tensor networks alone \cite{haghshenas2022variational}. Performing ACK on quantum circuit tensor networks allows us to achieve the sampling overhead given in \cref{eq:gamma-pure-state}. Since the $\gamma$ factor leads to an upper bound for entropy, ACK is particularly well-suited for problems with heterogeneous entanglement distribution. In other words, we can study many-body systems with low entanglement boundaries between high-entanglement regions. High-entanglement regions are desired to justify the use of distributed quantum computing instead of classical tensor network simulations. Low-entanglement regions are necessary so we can perform ACK with reasonable sampling overhead.
	
	\begin{figure}[b!]
		\centerline{\includegraphics[width=0.8\columnwidth]{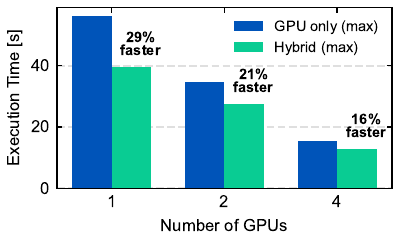}}
		\caption{ 
			Comparison between a GPU-only execution and the hybrid GPU-CPU approach to simulate sub-circuits (a total of 4000) resulting from partitioning a 30-qubit circuit into 10 and 20 qubits. Experiments were carried out on the Nvidia GH200 superchip.}
		\label{fig3}
	\end{figure}
	
	\begin{figure*}[t]
		\includegraphics{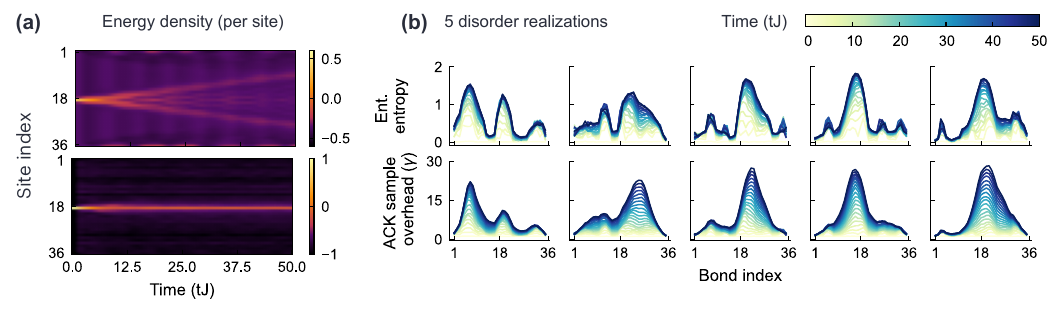}
		\caption{(a) Dynamics of an energy density for a clean (top) and  disordered (bottom) mixed-field Ising model with disorder in the longitudinal field. An initial energy perturbation diffuses in the clean model, whereas it gets localized in the disordered case. We used $J=1$, $h_z=1.01$, and $\Delta t = 0.25$. $h_x$ was set to 1 for the clean model whereas chosen randomly from $[-W,W]$, $W=2.00$, with a uniform probability for the disordered model. (b) Dynamics of entanglement entropy for 5 random disorder instances are shown in the top panel. In the bottom panel, we show the theoretical lower bound for ACK sample complexity, $\gamma$, which closely follows the entanglement entropy in the top panel. Since the total number of samples required is $O(\gamma^2/\epsilon^2)$, several orders of magnitude advantage can be achieved for simulating non-equilibrium dynamics up to $tJ=50$.
		}
		\label{fig4}
	\end{figure*}
	
	We present our algorithm in \cref{Fig2}. Input to the ACK algorithm is a tensor network representation of a quantum state, e.g., an MPS. The tensor network is partitioned according to the number of QPUs available, truncating cut bond dimensions. In the inner loop, for each partition, two-qubit unitary gates are variationally optimized to find a quantum circuit tensor network that effectively represents the tensor network partition given a specified circuit layout. Here, we use staircase circuits with order $M$. Unless otherwise specified, the numerical studies in this work use $M=3$, as it was found to provide an effective balance of accuracy and runtime. Note that the inner loop is trivially parallelizable, as each partition is independent. Once the sub-circuits are optimized, the outer loop constructs partial entanglement entropy heatmaps of the full system (missing measures between partitions), and recommends a new set of partitions at lower entanglement cut locations. The outer loop continues until the minimal entanglement cut locations are found, given the partial heatmaps available. Note that, because the system is not translationally invariant, entropy heatmaps are not uniform and not all cuts are equal in terms of sampling overhead. Once final partitions are determined, the two-qubit gates that cross partitions (the gates to be knit) are assigned from a separate iteration where partition locations are at least two qubits away. Finally, the optimized partitions and gates to be knit are passed to the circuit knitting subroutine for QPU simulation and observable reconstruction via QPD, as described in \cref{sec:circuit_knitting}. An algorithmic description of ACK is provided in \ref{sec:pseudocode}. 
	
	Note that the ACK algorithm as shown in \cref{Fig2} is general in that it assumes access only to QPUs that are smaller than the total size of the system to be simulated, and is intended for systems that are intractable for tensor networks without partitioning, as may be true for time-evolution circuits. A simplified version of the algorithm can be used for state preparation, where entanglement heatmaps are taken directly from the input tensor network which can describe the entire initial state, as we employ for the 2D results in \cref{sec:time_evo_2d}. In either case, the governing notion is to use entanglement measures to guide partitioning.

	\subsection{Parallel simulation and implementation}
	
	Classical simulation of quantum circuit execution is an essential for the development and benchmarking of quantum computing, especially as quantum devices are limited by size, noise, and availability. Moreover, high-performance classical simulations are critical for exploring the scalability of quantum algorithms, as they allow analyzing the computational requirements of algorithms on larger quantum systems before deploying them on physical quantum hardware. While the ACK algorithm is designed to enable efficient circuit knitting execution on real quantum devices, in this work we use high-performance classical simulation to implement, demonstrate, and test the algorithm.

	The ACK algorithm allows for multiple levels of parallelism and thus an efficient implementation for large-scale testing. Distributed memory parallelism for the simulation of sub-circuits in the the inner loop (left side of \cref{Fig2}) is accomplished by the message passing interface (MPI) via mpi4py \cite{mpi4py}, while CuPy~\cite{nishino2017cupy} has been used wherever applicable to replace NumPy for GPU execution, such as for the optimization of circuits against the target MPS and the generation of the initial MPS data. The resemblance of CuPy to NumPy significantly facilitated the porting process compared to a re-implementation. Furthermore, CuPy is not limited to CUDA devices and can also be used on AMD GPUs \cite{cupy-amd}. We conducted our experiments on an HPE-Cray EX system featuring four GH200 superchips per node. For circuit knitting execution (right side of \cref{Fig2}) we rely on a state-of-the-art toolkit (qiskit-addon-cutting) provided by Qiskit~\cite{qiskit-ckt} (CKT). Reference calculations as well as simulations of sub-circuits generated by the CKT have been executed with the Qiskit AER simulator using the MPS backend on the Grace CPU. 
	It should be noted that although GPUs have demonstrated substantial performance gains in simulating quantum circuits compared to CPUs \cite{bayraktar2023cuquantum, qiskit-aer, cirq-sim, van2023qclab++, jones2019quest, zhong2025scalable, oumarou2021fast}, no comparable effort has been made for specialized techniques like circuit knitting. 

	To better understand the potential of GPU simulations for circuit knitting, we conducted a separate investigation based on the assumption that the simulation of a large number of sub-circuits with potentially very different widths would benefit substantially from an guided load balancing across a hybrid CPU/GPU architecture. While a large relative overhead attributed to data transfer and reshaping can be expected from small circuits simulated on GPUs, larger circuits can hide this latency more efficiently. Thus, it would make sense to execute smaller circuits on the CPU instead, where the overhead can be avoided but the computation is still feasible. To demonstrate a potential advantage, we partition a representative 30-qubit circuit into sub-circuits of width 20 and 10 and distribute accordingly on GPUs and CPUs for hybrid execution. These circuits are derived from the Trotterized time evolution by a Hamiltonian such as \cref{eq:ham_long_disorder} but the results can be generalized as long as the circuit under consideration is similarly Trotterized.  \cref{fig3} shows a comparison between a GPU-only execution vs. the hybrid approach, giving a relative improvement of 15\% to 30\% while maintaining a low variance across MPI ranks. For these experiments, we ported the exact sampler from Qiskit for statevector simulations to GPUs using CuPy. In a similar way, the Qiskit AER's TensorNet backend could be considered. Furthermore, hybrid execution on CPUs and GPUs could be overlapped by making use of async methods to further reduce execution time.

	\section{Numerical studies}
	\label{sec:numerical_studies}
	
	\subsection{Time evolution in one-dimensional systems}
	\label{sec:time_evo_1d}
	
	\begin{figure*}[t]
		\centerline{\includegraphics[width=1.8\columnwidth]{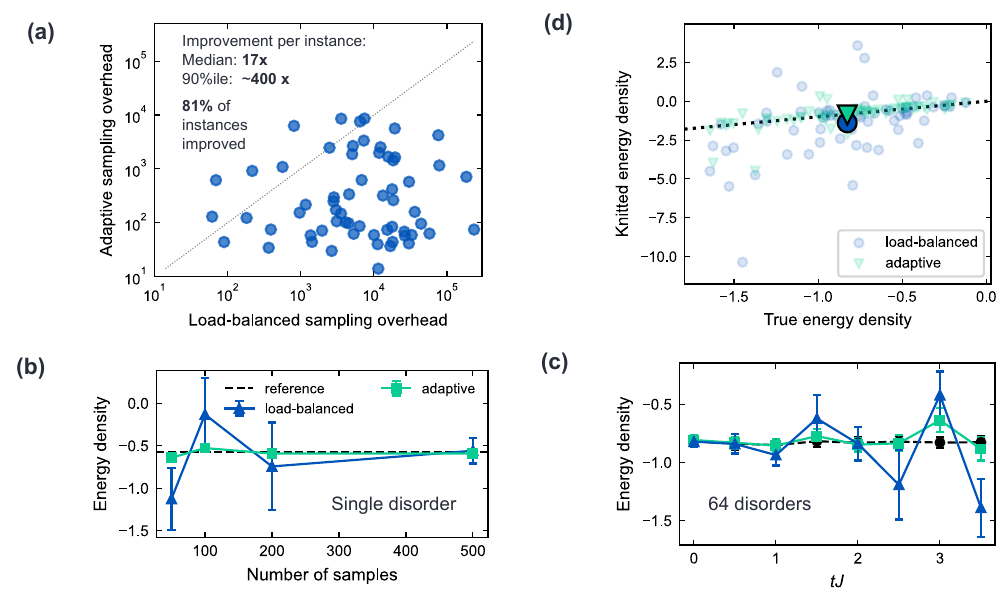}}
		\caption{ 
			Analysis of 40-qubit ACK simulations for time-evolution circuits at $tJ=\{0.0, 0.5, ...,3.5\}$ of the Hamiltonian given in \cref{eq:time_evo_hamiltonian} with 64 disorder instances. Here, J is the median among the randomly distributed $J_{j,j+1}$s. The observable considered is the energy density from \cref{eq:time_evo_edens} at site $j=13$. Values are compared for a load-balanced cut close to site 20 with a cut guided by entanglement (adaptive).
			(a) Comparison of circuit knitting sampling overheads at $tJ=3.5$. 
			(b) For a single illustrative disorder instance, the convergence of the energy density as a function of the number of circuit knitting samples, where 10 QPD repetitions were used to estimate the standard deviation.
			(c) Time evolution of the energy density (averaged over 64 disorder instances) computed with the load-balanced and adaptive knitting strategies (500 circuit knitting samples) compared to the true value from TEBD simulation.
			(d) Comparison of the knitted (load-balanced or adaptive) energy densities at $tJ=3.5$ using a fixed number of samples (500) compared to the true value computed with the Qiskit MPS simulator. The disorder-averaged energy densities are highlighted by the two bold symbols and the dotted line indicates the true energy density.}
		\label{fig5}
	\end{figure*}
	
	\begin{figure}[ht!]
		\centerline{\includegraphics[width=0.9\columnwidth]{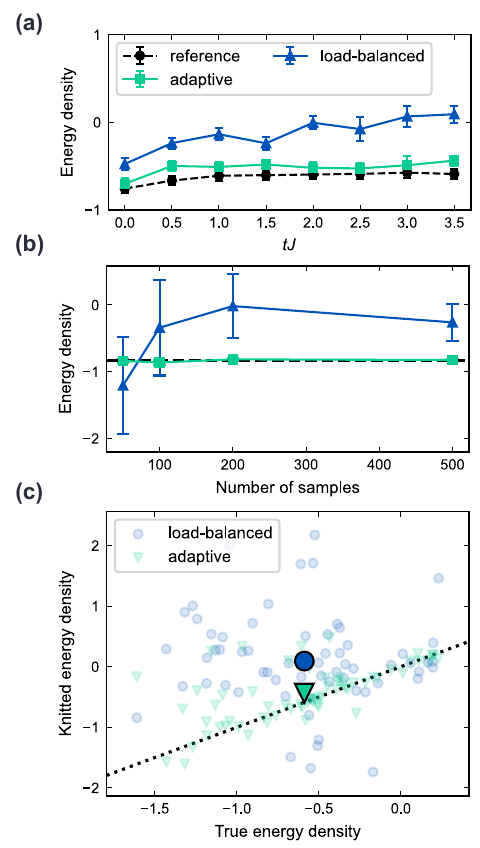}}
		\caption{Analysis of 40-qubit ACK simulations for time-evolution circuits at $tJ=\{0.0, 0.5, ...,3.5\}$ of the Hamiltonian given in \cref{eq:time_evo_hamiltonian} with 64 disorder instances. The observable considered is the energy density from \cref{eq:time_evo_edens} at site $j=20$. Values are compared for a load-balanced cut close to site 20 with a cut guided by entanglement (adaptive).
			(a) Time evolution of the energy density (averaged over disorder instances) computed with the load-balanced and adaptive knitting strategies (500 circuit knitting samples) compared to the true value from TEBD simulation. (b) For a single illustrative disorder instance, the convergence of the energy density as a function of the number of circuit knitting samples, where 10 QPD repetitions were used to estimate the standard deviation. (c) Comparison of the knitted (load-balanced or adaptive) energy densities at $tJ=3.5$ using a fixed number of samples (500) compared to the true value computed with the Qiskit MPS simulator. The disorder-averaged energy densities are highlighted by the two bold symbols; 
		}
		\label{fig6}
	\end{figure}
	
	In this section, we investigate the time evolution of disordered quantum many‑body systems in one and two spatial dimensions using adaptive circuit knitting (ACK). Before applying the ACK algorithm introduced in \cref{Fig2}, we first develop intuition for why disordered many‑body systems naturally lend themselves to distributed simulation strategies such as ACK. To this end, we study the mixed‑field Ising model with disorder in the longitudinal field. In one dimension, the Hamiltonian is
	\begin{equation}
		H=-\sum_{j=1}^{N-1}{\sigma_j^x \sigma_{j+1}^x} - g\sum_{j=1}^N{\sigma_j^z} - \sum_{j=1}^N{h_j \,\sigma_j^x}\, ,
		\label{eq:ham_long_disorder}
	\end{equation}
	
	\noindent where $\sigma^x$ and $\sigma^z$ are the Pauli $X$ and $Z$ matrices, respectively, $j$ indexes the qubits on the open chain, and $g$ and $h_j$s are real-valued parameters denoting the strength of the transverse and longitudinal fields, respectively. The disordered longitudinal field $h_j$ is chosen uniformly from $[-W, W]$. Without the longitudinal field, the model is integrable via mapping to free-fermions, whereas the longitudinal field breaks the integrability. Time-evolution under this non-integrable Hamiltonian generates quantum states with volume-law entanglement \cite{MariCarmen2011NonIntegrable}. This in turn means that such dynamics require exponential runtime and memory to simulate on classical computers. However, the introduction of disorder to the longitudinal field can fundamentally change the physics of the model from generating volume-law entanglement to area-law entanglement \cite{basko2006mbl,nandkishore2015mbl, kjall2014isingmbl,abanin2019colloquium, sierant2023stability}. Since such disordered systems do not have translation-invariance and therefore naturally host a heterogeneous entanglement structure, they are ideal candidates to study the efficacy of ACK techniques.
	
	We evaluate the success of ACK by reconstructing the energy density observable. Energy density per site for the above model is defined as
	\begin{equation}
		H_j=- \frac{1}{|N(j)|}\left( \sum_{k \in N(j)}{\sigma_j^x \sigma_{k}^x} \right)\, - g \, \sigma_j^z - {h_j \, \sigma_j^x}  
		\label{eq:state_prep_edens}
	\end{equation} 
	where $N(j)$ is the set of neighbors of site $j$. Because the total energy is conserved during unitary evolution, monitoring the spreading of an initially localized energy profile provides a clear diagnostic of localization. We prepare a low‑energy “bump” by aligning all spins along the direction $\ket{\hat{v}_j} \propto g \hat{z} + h_j \hat{x}$, except for the center spin, which is aligned along $\ket{-v_j}$. The resulting dynamics are shown in \cref{fig4}(a). In the clean case with no disorder in the longitudinal field ($g=1.01, h_j=1$), the energy bump diffuses away from the center, spreading uniformly across the system at late times. In contrast, the energy bump remains localized at the center in the disordered case as shown in the bottom panel, signaling a lack of energy transport.
	
	The propagation of conserved quantities is closely related to the propagation of quantum information \cite{abanin2019colloquium}. In the clean system, diffusive spreading of energy is accompanied by ballistic growth of entanglement entropy \cite{kimhuse2013ballisticspreading}. In contrast, in the many‑body‑localized regime, the disorder‑averaged entanglement entropy grows only logarithmically in time \cite{serbyn2013universalslow}. Because the disorder strength $W=2$ places the system in the crossover regime between ergodic and localized behavior \cite{sierant2023stability}, we expect the average entanglement‑entropy growth to lie between these two limits. The top panel of \cref{fig4}(b) shows the entanglement dynamics following an unentangled energy‑bump initial state. Across all five disorder realizations, we observe a heterogeneous entanglement profile, with regions of high entanglement separated by low‑entanglement boundaries that can be used to identify entanglement‑aware adaptive cuts. The bottom panel shows that the optimal circuit‑knitting overhead, quantified by the $\gamma$ factor, closely follows the entanglement entropy, consistent with the inequality in \cref{eq:entropy_inequality}. Guided by these observations, we now proceed to implement the ACK algorithm.
	
	Motivated by quantum spin glass phenomenon which arises in disordered Ising model \cite{rieger1996griffiths,Pekker2014}, we now study the mixed-field Ising model with disorder in all terms (fully disordered) given by
	\begin{equation}
		H=-\sum_{\langle i, j \rangle}{J_{i,j}\sigma_i^x \sigma_{j}^x} - \sum_{j=1}^N{g_j\sigma_j^z} - \sum_{j=1}^N{h_j \sigma_j^x}\, 
		\label{eq:time_evo_hamiltonian}
	\end{equation}
	where $\sigma^z$ and $\sigma^x$ are the Pauli $Z$ and $X$ matrices, respectively, $\langle i,j \rangle$ denotes the nearest neighbors, and the \(\{J_{i,j},g_{j},h_{j}\}\)'s are real-valued parameters denoting the strength of the coupling term, transverse field, and longitudinal field, respectively. Parameters are chosen randomly from a uniform distribution on [-1, 1].

	Energy density per site for our model is defined as
	\begin{equation}
		H_j=- g_j \, \sigma_j^z - {h_j \sigma_j^x} - \frac{1}{|N(j)|}\sum_{k \in N(j)}{J_{jk} \sigma_j^x \sigma_{k}^x} \, 
		\label{eq:time_evo_edens}
	\end{equation}

	We evaluate the cost and accuracy of circuit knitting, comparing adaptive cut locations to load-balanced cut locations for calculating the energy density \cref{eq:time_evo_edens} under the time evolution of the Hamiltonian \cref{eq:time_evo_hamiltonian} for 40-qubit systems. For each time in $tJ=\{0.0,0.5,\ldots,3.5\}$, 64 disorder instances are considered and the energy density is evaluated at two sites on the chain, $j=13$ and $j=20$. Reference MPS data for each of the time points and disorder instances is generated from a time-evolving block-decimation (TEBD) simulation \cite{dmrg_2011_schollwock, lin2021real}. We show the analysis for site $j=13$ in \cref{fig5} and for site $j=20$ in  \cref{fig6}.
	
	When site $j=13$ is considered, the energy density is reproduced by both methods, though the adaptive approach approximates the true value more accurately at a fixed number of samples, as can be seen in \cref{fig5}(b), (c), and (d). Sampling overheads resulting from the adaptive strategy are lower than load-balanced by a median factor of 17 (for 90th percentile, a factor of 459), with 81.2\% of disorder instances improved, shown as falling below the diagonal---see \cref{fig5}(a). We also note that for site $j=13$ the load-balanced approach becomes increasingly inaccurate for larger times compared to the adaptive strategy---see \cref{fig5}(c). These results indicate that, given a fixed number of samples (i.e., a fixed amount of QPU resources), the adaptive strategy will result in more accurate knitted observables. The sampling advantage of the adaptive strategy comes from the lower variance of the QPD. This can be quantified by a lower gamma factor corresponding to a low-entanglement cut, as we saw in \cref{fig4}. Although the gamma factor reported in this figure was for cutting the whole state, we can expect a similar trend in the total gamma factor for a variationally optimized quantum circuit.
	
	\begin{figure}[t]
		\centerline{\includegraphics[width=0.85\columnwidth]{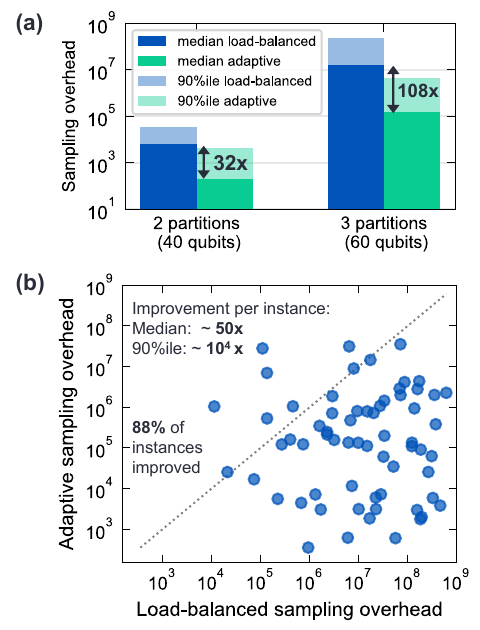}}
		\caption{ Analysis of 60-qubit ACK simulations for time-evolution circuits at $tJ=3.5$ of the Hamiltonian given in \cref{eq:time_evo_hamiltonian} with 64 disorder instances. The observable considered is the energy density from \cref{eq:time_evo_edens} at the middle of the chain. For 60-qubit simulations, ACK cuts between two sets of qubits result in 3 partitions. (a) Median and 90th percentile circuit knitting sampling overheads for load-balanced and adaptive cuts on 40-qubit and 60-qubit systems. Reduction in overhead due to the adaptive scheme increases as more gates are cut (from 2 to 3 partitions), as shown in the widening gap for median overheads. (b) Comparison of circuit knitting sampling overheads for 60-qubit systems with 3 partitions (with order $M=3$, a total of six two-qubit unitaries are cut). 
		}
		\label{fig7}
	\end{figure}
	
	\begin{figure}[ht!]
		\centerline{\includegraphics[width=0.95\columnwidth]{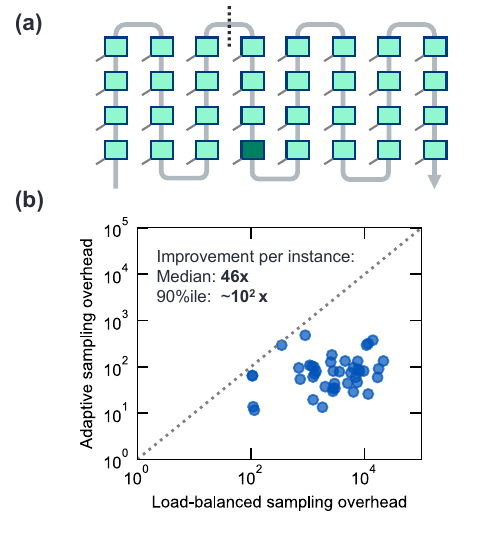}}
		\caption{ 32-qubit ACK simulations for time-evolution circuits at {$t=4.0$} of the 2D Hamiltonian given in \cref{eq:time_evo_hamiltonian} with 43 disorder instances. (a) Diagaram of the snake topology used to simulate the 4x8 lattice with an MPS. An example cut is shown with a dotted line. (b) Comparison of circuit knitting sampling overheads for reconstructing the energy density observable on the lattice site at the edge in the middle of the chain, as shown dark green in (a). Adaptive cuts show improvement from all instances with a median improvement or reduction in overhead per disorder instance of a factor of 46 (for 90th percentile, close to a factor of 121 improvement). }
		\label{fig8}
	\end{figure}
	
	Similarly, in \cref{fig6} we show the energy density at the center of the chain, site $j=20$, demonstrating the adaptive strategy far outperforms the load-balanced strategy in both accuracy and robustness. In this case, the observables are measured near the load-balanced cut for all disorder instances. Hence, the load-balanced circuit knitting is at an inherent disadvantage since two-qubit gates from a separate iteration, as explained in \cref{sec:ack}, are directly acting on the qubits of interest. This highlights yet another benefit of the adaptive approach: because cut locations are varied on an instance-by-instance basis, the knitting procedure is less sensitive to specific observables. Furthermore, we note that the adaptive data points relatively away from the diagonal in \cref{fig6}(c) are the ones from later times. Similarly to \cref{fig5}(c), this also suggests that given a target accuracy, we should dynamically choose the number of samples depending on the entanglement entropy, and hence the 1-norm of the QPD, the gamma factor.
	
	To investigate how ACK can scale for larger systems, we simulate the same Hamiltonian as described above for 60-qubit disorder instances, in these cases cutting in two locations to produce three partitions (a total of six two-qubit unitary gate cuts). The analysis is summarized in \cref{fig7}. \cref{fig7}(a) compares the median and 90th percentile sampling overheads for the 40-qubit simulations (described above) with the 60-qubit simulations. The adaptive strategy has lower overheads than the load-balanced by one to two orders of magnitude, and we note that, while the sampling overheads of course increase with the number of gates cut, the gap between the adaptive and load-balanced approaches widens when more gates are cut. \cref{fig7}(b) compares sampling overheads across all disorder instances for the 60-qubit system, and shows that sampling overheads for the adaptive strategy are smaller than load-balanced by a median factor of 51 (90th percentile factor of 10167), with 87.5\% of instances showing improvement. Thus, our scaling study indicates that ACK can achieve even more savings for larger systems, where multiple cuts become necessary.

	\subsection{Time evolution in two-dimensional systems}
	\label{sec:time_evo_2d}
	While 1D spin systems are a sensible place to begin evaluating the power of quantum circuit tensor networks and methods like ACK, accurately simulating 2D systems is where many classical techniques break down and therefore a promising application area for quantum algorithms. For an initial evaluation of how the ACK method might perform on 2D systems, we extended our 1D implementation to accommodate 2D spin lattices via a MPS. As shown in \cref{fig8}(a), the 2D lattice is mapped to a 1D chain via a snake-like transversal that alternates direction on successive rows. Bond dimensions are expected to be much larger than in the 1D case in order to carry entanglement horizontally across the system. The dashed line in \cref{fig8}(a) indicates a representative circuit cut location along the snake ordering.

	We again study the mixed-field Ising model from \cref{eq:time_evo_hamiltonian}, defined in two dimensions. The initial energy-bump state, similar to the one described for 1D, is constructed by locally inverting the effective field orientation at an edge-middle site of the lattice, which is shown in dark green in \cref{fig8}(a). We observe qualitatively similar behavior to 1D in the fully disordered 
	4 × 8 (32-qubit) mixed-field Ising model with a localized energy-bump initialization. As expected, the introduction of two-dimensional connectivity significantly enhances the overall entanglement growth along the snake-mapped MPS chain. Meanwhile, as the snake-mapping results in an effectively quasi-1D chain with 8 qudits, there is less variation in entanglement compared to 1D. Unlike the 1D case, we do not perform full circuit-knitting simulations for the 2D case; instead, we focus on analyzing the expected sampling overhead based on variationally optimized circuits that maximize the fidelity with respect to the target MPS. We note that the fidelities were in the 0.95-1 range, with a median fidelity of 0.97, as opposed to a median of 0.99 in the 1D case. The adaptive circuit knitting strategy consistently shows a lower overhead compared to load-balanced cutting. The resulting median improvement is $46$x with all of the disorder instances improved by our ACK algorithm using the bipartite von Neumann entropy for choosing the cuts.
	It is not surprising that all adaptive cuts lowers the sampling overhead for all the datapoints since we have an exact entanglement heatmap for the entire system as opposed to iteratively choosing the cut in the outer loop of the ACK algorithm in \cref{Fig2}. The observed reduction in sampling overhead shows the snake MPS ansatz to be effective in capturing the entanglement structure of this two-dimensional quantum system. However, the bond dimension even for an area-law state grows exponentially in the length of the shorter dimension, so this equivalence is limited to quasi-1D systems. Thus, for the application of ACK to a fully 2D system, we will require 2D tensor networks such as PEPS.

	\section{Discussion}
	\label{sec:discussion}
	In this work, we have primarily focused on matrix product states (MPS) as a basis for distributing quantum workloads. This choice served as an initial demonstration of both the feasibility of distributed quantum simulation and the advantages of using an adaptive strategy to determine circuit cuts. While MPS‑based circuits provide a convenient starting point, the ACK framework itself is not restricted to this ansatz. In future work, we will explore how ACK can be applied to more general tensor‑network families, including MERA, PEPS, and related structures.
	
	Although the algorithm shown in \cref{Fig2} is intended for scenarios in which the appropriate tensor‑network representation of the target state is not known a priori, the application of ACK to preparing a known MPS is considerably more direct. Distributed state preparation will be a critical subroutine in distributed quantum simulation pipelines, where a future distributed algorithm for the time‑evolution circuit is applied to a low‑energy state \cite{gyawali2026ack}.
	
	For our initial application of ACK, we considered the fully disordered Ising model. In this setting, the inner loop of the algorithm in \cref{Fig2} patches boundary qubits residing on multiple QPUs using two-qubit gates that are obtained from a similar optimization but with different cut location. The agreement between the exact vs. circuit-knitted expectation values indicate that this strategy works reasonably well for the fully-disordered model. However, for systems which are not strongly disordered, it is a better strategy to simultaneously optimize the two-qubit gates at the boundary. We will investigate this in our future work.
	
	A central step in locating optimal cuts for ACK is characterizing the entanglement across prospective circuit boundaries. However, directly computing the exact sampling overhead $\gamma$, or equivalently the von Neumann entropy, exhibits a sample complexity that scales prohibitively with subsystem size. By contrast, the second Rényi entropy is more tractable and can be efficiently estimated on hardware \cite{brydges_2019_probing_renyi, huang_2020_predicting}. Intuitively, second‑R\'enyi measurements are easier because they require only two copies of the state, whereas measuring $\gamma$ or the von Neumann entropy would require access to infinitely many; hence, full tomography is necessary. In our analysis of entanglement entropy in \cref{fig4}, we observed that for the time‑translation‑invariant circuits studied, the characteristic entanglement structure starts to emerge at early times. This implies that early‑time entanglement dynamics, accessible through tensor‑network simulations, can serve as a practical guide for identifying promising circuit locations for cutting the circuit. Furthermore, machine-learning models which learn to predict the entanglement dynamics for a given disorder realization without simulating the full system would also be an interesting future direction to explore.
	
	While disordered quantum systems provide a natural testbed for ACK, the scope of the method extends well beyond this setting. A notable example is the quantum Fourier transform (QFT), which appears in central quantum algorithms such as Shor’s factoring and quantum phase estimation. The QFT is known to exhibit low (constant‑in‑system‑size) operator entanglement \cite{aharonov_2007_quantum_fft_classically_simulated, chen_2023_quantum_fourier_transform}, making it particularly suitable for efficient ACK‑based simulation.  More broadly, ACK offers practical value in distributed quantum computing architectures connected through quantum interconnects, where reducing the consumption of shared entangled resources, such as Bell pairs, is essential for achieving scalable performance.

	Likewise, entanglement-based partitioning strategies—of which ACK is a concrete realization—play a pivotal role within an HPC–QC full‑stack framework \cite{xin2025hpcqc}. In this architecture, such strategies operate at the hypervisor layer, dynamically identifying and partitioning large quantum circuits into sub-circuits that can be dispatched across multiple quantum devices and classical simulators. They can be implemented for NISQ, partially, and fully fault-tolerant architectures. By leveraging entanglement structure as a systems-level guiding principle, the framework enables more informed workload decomposition, resource allocation, and sampling cost management. ACK serves as a practical instantiation of these ideas, demonstrating how entanglement-aware circuit decomposition can be integrated with distributed runtime orchestration. Together, these capabilities enable scalable distributed hybrid execution and provide a unified pathway toward utility-scale quantum–classical computing.
	
	\section{Conclusions}
	\label{sec:conclusion}
	By leveraging the fact that low‑entanglement cuts yield a compressed Schmidt decomposition, we proposed adaptive circuit knitting algorithm that produces a correspondingly compressed quasiprobability decomposition. This leads to a practical and scalable approach for distributing quantum circuits across multiple QPUs, with disordered quantum systems as a natural testbed. For the fully disordered Ising model, we showed that the exponential sampling complexity of straightforward circuit knitting can be substantially reduced, up to four orders in magnitude, by cutting circuits at low‑entanglement boundaries. The same strategy facilitates efficient distributed classical simulation across heterogeneous CPU and GPU resources. As quantum hardware and classical accelerators continue to scale, such entanglement‑aware distributed approaches will play an increasingly central role in the roadmap toward a quantum supercomputer.
	
	\section*{Acknowledgments}
	
	We gratefully acknowledge our collaborators at NVIDIA, Pooja Rao and Yuri Alexeev, for insightful discussions on the ACK algorithm and parallel, GPU-accelerated implementation. We thank Fazilat Fatima (HPE) for help with baseline simulation studies, and Michael Ferguson (HPE) for review of the manuscript. We acknowledge support from DARPA’s Quantum Benchmarking Initiative (QBI), contract no HR00112590116.
	
	\section*{Data availability}
	
	Data used for generating the figures are available at Zenodo \cite{johnson_2026_distributed}. 
	
	\bibliography{main}
	
	\onecolumngrid
	\appendix
	\clearpage
	\section{ACK sampling overhead vs R\'enyi entropy}
	\label{sec:renyi_entropy}
	Here, we show that the sampling overhead $\gamma$ introduced in \cref{eq:gamma_factor} is the related to one of the R\'enyi entropies, a generalization of the von Neumann entropy. The R\'enyi entropy of order $\alpha$, where $0 < \alpha < \infty$ and $\alpha\neq1$, for the reduced density matrix $\rho$ is defined as
	\begin{equation}
		\label{eqn:renyi_definition}
		S_\alpha(\rho) = \frac{1}{1-\alpha} \log(\Tr(\rho^\alpha))
	\end{equation}
	The eigenvalues of $\rho$, say $\{p_0, p_1 ,\cdots, p_D\}$, are positive semi-definite hence form a probability distribution. The R\'enyi entropy can then be expressed as
	\begin{equation}
		S_\alpha(\rho) = \frac{1}{1-\alpha} \log(\sum_{j=1}^D p_j^\alpha)
	\end{equation}
	Taking limit $\alpha \to 1$ gives us the familiar von Neumann entropy. For the bipartite pure state in \cref{eq:bipartite_schmidt}, $p_j = \lambda_j^2$, where $\lambda_j$s are the Schmidt coefficients. $\alpha=1/2$ is of special interest for the ACK algorithm since
	\begin{align}
		S_{1/2}(\rho) = 2 \log(\sum_j^D p_j^{1/2}) &= 2 \log(\sum_j^D \lambda_j) \nonumber\\
		&= \log(\frac{\gamma+1}{2}).
	\end{align}
	In particular, this highlights that the ACK overhead is exponential in the R\'enyi entropy of the quantum state, \(S_{1/2}(\rho)\).
	
	Now, what is the relationship between the von Neumann entropy $S_{vN}(\rho)$ and the sampling overhead? Noting $S_\alpha$ is a non-increasing function of $\alpha$ \cite{muller_2014_on_renyi_entropies}, 
	\begin{equation}
		S_{1/2}(\rho) \geq S_{vN} (\rho) \geq S_2 (\rho),
	\end{equation}
	where $S_2(\rho)$ is the second R\'enyi entropy. Although the von Neumann entropy is closer to the half-R\'enyi entropy than the second-R\'enyi entropy, it is experimentally more feasible to estimate the second-R\'enyi entropy via randomized measurement protocols \cite{brydges_2019_probing_renyi, huang_2020_predicting}. In fact, accurate measurement of second-Renyi entropy for up to $16$ qubits has already been demonstrated on superconducting quantum processor \cite{gyawali_dfl}. Likewise, R\'enyi entropies are also easily accessible in quantum Monte Carlo simulations \cite{PhysRevLett.104.157201}.
	
	\clearpage
	\section{Adaptive circuit knitting algorithm}
	\label{sec:pseudocode}
	
	\begin{algorithm}[H]
		\SetAlgoLined
		\SetInd{1em}{1em}
		\caption{ \textsc{ACK from MPS}}\label{alg:ACK_MPS}
		\SetKwInput{Input}{Input}
		\SetKwInput{Output}{Output}
		\Input{$\ket{\psi_\text{MPS} (\{\chi\})}$ of length $N$ (number of qubits), number of QPUs $p$, quantum circuit order $M$}
		\Output{Optimized quantum circuit $\ket{\psi^M_\text{QC}}$ representing $\ket{\psi_\text{MPS} (\{\chi\})}$, recommended set of cut locations $K=\{k_1,...,k_{p-1}\}$ (cutting in space across full circuit).}
		
		\textbf{Initialize $K$} (e.g., from load-balanced or random cuts)
		
		\While{$K$ not previously seen}{
			
			\textbf{Partition} $\ket{\psi_\text{MPS} (\{\chi\})}$ into $p$ MPS states at $K$, truncating cut bond dims to $\chi=1$. 
			
			\textbf{Initialize $S$} (length $N-1$) for entanglement entropy between neighboring qubits
			
			\For{each partition $\ket{\psi_{\text{MPS,}p}}$ (in parallel)}{
				
				\textbf{Optimize gates in partition}: Find circuit unitaries $\{U(\theta)\}$ of order $M$ that maximize $\braket{\psi_{\text{QC,}p}^M (\{U(\theta)\}) | \psi_{\text{MPS,}p}}$
				
				\textbf{Compute ent. entropy} between qubits $i,i+1$ in partition and store in $S[i]$
				
			}
			
			\textbf{Update $K$} to lower ent. locations given $S$ (partial heatmap, missing between $\ket{\psi_{\text{MPS,}p}}$)
			
		}
		\textbf{Assemble full circuit}: Concatenate $\ket{\psi_{\text{QC,}p}^M}$ and $U(\theta^M_k)$ at $K$ (stored from intermed. or extra iteration) to form $\ket{\psi^M_\text{QC}}$.
		
		\textbf{Run circuit knitting}: Given $\ket{\psi^M_\text{QC}}$ and $K$, compute desired observable
		using $p$ QPUs.
		
	\end{algorithm}
\end{document}